\documentclass[aps,superscriptaddress,showpacs,preprint]{revtex4} 
\usepackage{graphicx,epsfig,pstricks}
\usepackage{amssymb}
\usepackage{amsmath}
\usepackage{dsfont}

\begin{document}

\title{Pion dissociation and Levinson's theorem in hot PNJL quark matter}

\author{A.~Wergieluk} \email[E-mail: ]{agnieszka@wergieluk.com}
\affiliation{Institute for Theoretical Physics, University of Wroc{\l}aw, 
50-204 Wroc{\l}aw, Poland} 

\author{D.~Blaschke} 
\affiliation{Institute for Theoretical Physics, University of Wroc{\l}aw, 
50-204 Wroc{\l}aw, Poland} 
\affiliation{Bogoliubov Laboratory for Theoretical Physics, JINR Dubna, 
141980 Dubna, Russia}
\affiliation{Fakult\"at f\"ur Physik, Universit\"at Bielefeld, 
33615 Bielefeld, Germany}

\author{Yu.~L.~Kalinovsky} 
\affiliation{Laboratory for Information Technologies, JINR Dubna, 
141980 Dubna, Russia}

\author{A.~V.~Friesen} 
\affiliation{Bogoliubov Laboratory for Theoretical Physics, JINR Dubna, 
141980 Dubna, Russia}

\date{\today}

\begin{abstract}
Pion dissociation by the Mott effect in quark plasma is described within the 
generalized Beth-Uhlenbeck approach on the basis of the PNJL model which 
allows for a unified description of bound, resonant and scattering states. 
As a first approximation, we utilize the Breit-Wigner ansatz for the spectral 
function and clarify its relation to the complex mass pole solution of the 
pion Bethe-Salpeter equation. 
Application of the Levinson theorem proves that describing the pion Mott 
dissociation solely by means of spectral broadening of the pion bound state 
beyond $T_{\rm Mott}$ leaves out a significant aspect. 
Thus we acknowledge the importance of the continuum of scattering states and 
show its role for the thermodynamics of pion dissociation.
\end{abstract}
\preprint{JINR Dubna Report No. E2-2013-19}
\maketitle

\section{Introduction}
We investigate the thermodynamics of mesonic bound states in hot medium consisting of a nonideal quark plasma with correlations in the pion and sigma meson interaction channels. A special emphasis is put on a correct description of the bound states' dissociation in the vicinity and beyond the Mott temperature. To this end we utilize the Polyakov-loop-extended Nambu--Jona-Lasinio model at finite temperature. 

The model has two order parameters: the chiral condensate,      
determining the value of the dynamically generated quark mass $m(T)$
related to the chiral symmetry breaking/restoration transition, and the
Polyakov loop variable $\Phi(T)$ which is an order parameter for
deconfinement. Their values are obtained from a self-consistent solutions of the coupled gap equations for these parameters which correspond to the
location of the minimum of the thermodynamical potential $\Omega(m,\Phi;T)$
in the mean-field approximation (MFA). Going beyond the MFA, we evaluate the contributions from pion and sigma meson fluctuations within the Gaussian approximation to the path-integral representation of the thermodynamic potential. Above the Mott temperature one observes the spectral broadening of bound states leading to the appearance of complex mass poles in the mesonic propagators. A proper analysis leads to coupled Bethe-Salpeter equations from which meson masses and corresponding spectral widths are obtained. Next, the equation of state for the quark-meson system is obtained in form of a generalized Beth-Uhlenbeck equation \cite{Beth-Uhl:456} which describes the effects that chiral symmetry restoration and deconfinement have on the contributions from pions and sigma mesons and allows for an adequate description of bound, resonant and scattering states on equal footing. The spectral functions in the mesonic channels are evaluated from the complex-valued polarization loop integrals. An examination of the Levinson theorem \cite{Dashen:1969ep} proves that in order to correctly describe the Mott effect, the continuum of mesonic correlations (scattering states) has to be taken into the account. As a result of this analysis, we obtain a description of pion dissociation where the effect of the vanishing bound state is exactly compensated by the occurrence of a resonance in the continuum of scattering states. 

The present work improves on previous works within the NJL model \cite{Hufner:1994ma,Zhuang:1994dw}, where unphysical quark degrees of freedom appeared in the hadronic phase due to the lack of the coupling to the Polyakov-loop and the Stefan-Boltzmann limit was not obtained due to a misplaced momentum cutoff. We also improve the recent work \cite{Rossner:2007ik} by properly discussing the role of Levinson's theorem in the PNJL model approach to the generalized Beth-Uhlenbeck equation of state for the quark-meson plasma, see also recent developments in this direction in Refs.~\cite{Yamazaki:2012ux,Blaschke:2013zaa}.
The model presented joins both exact limits of finite-temperature QCD - the pion gas at low temperatures and the quark-gluon Stefan-Boltzmann limit at high temperatures - within a microscopic chiral quark model approach.

\section{Quarks and light mesons in the PNJL model}

The definition of the Nambu--Jona-Lasinio model \cite{Nambu:1961tp}-\cite{Buballa:2003qv} with Polyakov loop \cite{Meisinger:2001cq}-\cite{Hansen:2006ee} is given in the Appendix \ref{appA123}. Here we start from the expansion of the thermodynamic potential
\begin{eqnarray}
\Omega \left(T, \mu \right) = - \frac{T}{V} \ln {\mathcal Z} \left[T, V, \mu \right]
\end{eqnarray}
around homogenous mean field values, which leads to the decomposition of the auxiliary fields $\sigma'$ and $\vec{\pi}'$ into their mean field and fluctuation  parts
\begin{eqnarray}
\sigma' = \sigma_{MF} + \sigma, && \vec{\pi}' = \vec{\pi} \hspace{15pt}  (\vec{\pi}_{MF} = 0)~.
\label{splitting_pion}
\end{eqnarray}
Utilizing the decomposed fields results in the factorization of the partition function into the respective parts describing mean field and fluctuation  contributions:
\begin{eqnarray}
{\mathcal Z}_{MF} [T, V, \mu] =  \exp \left\{ - \frac{V}{T} \bigg(\frac{\sigma_{MF}^2}{4 G_S} + U(\Phi, \overline{\Phi}; T) \bigg) + {\rm Tr} \ln \Big[ \beta S_{MF}^{-1} [m]\Big] \right\}~,
\end{eqnarray}
\begin{eqnarray}
{\mathcal Z}_{FL} [T,V,\mu] &=& \int {\mathcal D}\sigma {\mathcal D}\vec{\pi} \exp \Bigg\{ - \bigg[ \int_0^{\beta} d\tau \int_V d^3x \hspace{3pt} \frac{2\sigma \sigma_{MF} + \sigma^2 + \vec{\pi}^2}{4 G_S} \bigg] \nonumber\\
&&+  {\rm Tr} \ln \Big[ 1 -  S_{MF}[m]\left(\sigma + i \gamma_5 \vec{\tau} \vec{\pi} \right) \Big] \Bigg\}~.
\label{Z_fluctuations_01}
\end{eqnarray}
Here the mean field inverse propagator is
\begin{eqnarray}
S^{-1}_{MF} = \gamma^0(i \omega_n - \mu + A_0) - \vec{\gamma}\cdot\vec{p} - m_0 - \sigma_{MF} = \gamma^0(i \omega_n - \mu + A_0) - \vec{\gamma}\cdot\vec{p} - m~.
\label{PNJL_inverse_mean_field_quark_propagator}
\end{eqnarray}
The thermodynamic potential in the mean field approximation of the PNJL model is evaluated to be given by the following expression~\cite{Hansen:2006ee}
\begin{eqnarray}
&& \Omega_{MF} = \frac{\sigma_{MF}^2}{4G_S}  
+ U(\Phi, \overline{\Phi}; T) 
-  2N_f  \int \frac{d^3p}{(2\pi)^3} \left\{N_c  E_p  
+  T \bigg[  \ln N_\Phi^-(E_p)+ \ln N_\Phi^+(E_p)  \bigg]\right\}~,
\end{eqnarray}
where factors $N_f$, $N_c$ originate from performing the trace operation and are a consequence of isospin and color symmetry. The quark energy is given by $E_p=\sqrt{{\bf p}^2+m^2}$, $E_p^\mp$ are defined as $E_p^\mp = E_p\mp \mu$, and
\begin{eqnarray}
&& N^-_\Phi(E_p) = \left[ 1+3\left( \Phi +\overline{\Phi} e^{-\beta
E_p^-}\right) e^{-\beta E_p^-} + e^{-3\beta E_p^-}
\right]~, \\
&& N^+_\Phi(E_p) = \left[ 1+3\left( \overline{\Phi} + {\Phi} e^{-\beta
E_p^+}\right) e^{-\beta E_p^+} + e^{-3\beta E_p^+} \right]~.
\end{eqnarray}

In the mean field approximation of the PNJL model the values of the constituent
quark mass $m$ and the Polyakov-loop variable $\Phi$, along with its complex 
conjugate $\overline\Phi$, are obtained from the condition that the thermodynamic 
potential should be minimized with respect to these parameters, which is augmented by the stability conditions.
For $\mu = 0$, we have $\Phi = \overline{\Phi}$ and thus the minimizing conditions 
are given by
\begin{eqnarray}
	\frac{\partial \Omega_{MF}}{\partial \sigma_{MF}} = 0,  &&   \frac{\partial 	\Omega_{MF}}{\partial \Phi} = 0~.
\label{gap_equations}
\end{eqnarray}
These conditions are equivalent to a set of coupled gap 
equations~\cite{Ratti:2005jh,Hansen:2006ee}. For the mass gap equation we get
\begin{eqnarray}
m &=& m_0 +  4N_fN_c G_S  \int^{\Lambda} \frac{d^3p}{(2\pi)^3} ~ \frac{m}{E_p}  \bigg[1 - f^{-}_{\Phi}(E_p) - f^{+}_{\Phi}(E_p)\bigg]~, 
\label{masq}
\end{eqnarray}
where
\begin{eqnarray}
f^{\mp}_{\Phi}(E_p) = \big[ \Phi e^{-\beta(E_p \mp \mu)} + 2\Phi e^{-2\beta(E_p \mp \mu)} + e^{-3\beta(E_p \mp \mu)}\big]/N^{\mp}_{\Phi}(E_p)
\label{gen_fermi}
\end{eqnarray}
are the so-called generalized Fermi functions, characteristic for the PNJL model. One should note that if $\Phi \rightarrow 1$, the expression (\ref{gen_fermi}) reduces to the standard NJL model Fermi functions. For PNJL calculations we should additionally find the values of $\Phi$ from corresponding gap equation \cite{Hansen:2006ee} at given $T$ and $\mu$. 

In order to solve (\ref{masq}), a 
set of model parameters has to be determined: the cutoff parameter $\Lambda$, 
the current quark mass $m_0$ (in the chiral limit $m_0=0$) and the coupling
constant $G_S$. 
These parameters are fitted at $T = 0$ to reproduce physical quantities: the 
pion mass $M_\pi = 135$ MeV, the pion decay constant $F_\pi = 92.4$ MeV and 
the quark condensate $\langle \overline{q} q\rangle^{1/3}=-240.772$ MeV. 
The used parameters \cite{Grigorian:2006qe} are shown in Table~\ref{table2}.
\begin{figure}
\centerline{
\psfig{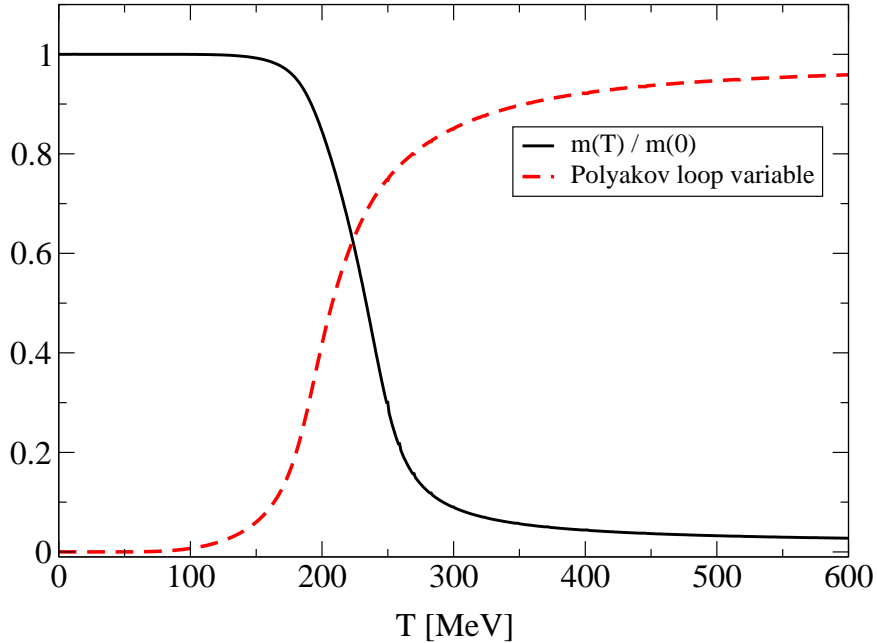}
}
\caption{ Temperature dependence of the quark masses $m(T)$ and Polyakov loop variable $\Phi(T)$ at $\mu$ = 0. Results for quark mass are scaled by $m(0) = 367.5$ MeV.}
\label{quark_mass_phi}
\end{figure}

\begin{table}
\caption{The set of model parameters reproducing observable quantities 
(in brackets) and $\langle \overline{q} q\rangle^{1/3}=-240.772$ MeV \cite{Grigorian:2006qe}:}
\begin{tabular}{|c|c|c|c|c|}\toprule
$m_0$ [MeV] & $\Lambda$ [MeV] & $G_S\Lambda^2$ & $F_\pi$ [MeV] & $M_\pi$ [MeV] \\ \colrule
5.495 & 602.472 & 2.318 & (92.4) & (135) \\
\botrule
\end{tabular}
\label{table2}
\end{table}

Since NJL-type models are non-renormalizable, it is necessary to introduce a 
regularization, e.g., by a cutoff $\Lambda$ in the momentum integration.
Following \cite{Hansen:2006ee}, in this study, in case of thermodynamic quantities, we use the three-dimensional
momentum cutoff for vacuum terms and extend the integration till infinity for finite temperatures. A comprehensive study of the differences between the two regularization procedures (with and without cutoff on the quark momentum states at finite temperature) has been performed in \cite{Costa:2009ae}.

Solutions of the gap equation (\ref{masq}) and the corresponding gap equation for the Polyakov loop variable $\Phi$ at zero chemical potential and nonzero $T$ are presented in Fig.~\ref{quark_mass_phi}. Above the critical temperature, which is equal $T_c = 237$ MeV in the chiral limit and $T_c = 251$ MeV for a finite current quark mass, one observes chiral symmetry restoration indicated by the rapid decrease of the constituent quark mass and the Polyakov loop variable becoming close to 1. The corresponding mean field contribution to the pressure, given by $P = - \Omega_{MF}$, is shown on Fig. \ref{pion_press_BW_phi_2} (scaled by a factor $T^4$) in section VI. One observes that in opposition to classical NJL models (e.g., \cite{Hufner:1994ma,Zhuang:1994dw}), quark degrees of freedom are suppresed below the critical temperature in the PNJL model. Moreover, above the critical temperature, the gluonic degrees of freedom are accounted for correctly.

The contribution to the thermodynamics stemming from the fluctuations described by (\ref{Z_fluctuations_01}) is evaluated in a scheme where we expand the logarithm up to the second (Gaussian) order according to
\begin{eqnarray}
\ln \left(1 - x \right) = - \sum_{k=1}^{\infty} \frac{x^k}{k} = -x - \frac{1}{2} x^2 + \dots, && |x| < 1~,
\end{eqnarray}
to get
\begin{eqnarray}
{\mathcal Z}^{(2)}_{FL} [T,V,\mu] &=& 
\int {\mathcal D}\sigma {\mathcal D}\vec{\pi} 
\exp \Bigg\{ - \bigg[ \int_0^{\beta} d\tau \int_V d^3x ~ 
\frac{\sigma^2 + \vec{\pi}^2}{4 G_S} \bigg] \nonumber\\
&&-  \frac{1}{2} {\rm Tr} 
\Big(S_{MF}[m] \Sigma[\sigma, \vec{\pi}] S_{MF}[m] \Sigma[\sigma, \vec{\pi}] 
\Big) \Bigg\}~,
\end{eqnarray}
where we have introduced 
$\Sigma [\sigma, \vec{\pi}] = \sigma + i \gamma_5 \vec{\tau} \vec{\pi}$.

Performing the calculation leads to the subsequent factorization of the thermodynamic potential into parts describing the contribution from mesonic correlations corresponding to $\sigma$ and $\vec{\pi}$ channels of interaction
\begin{eqnarray}
{\mathcal Z}^{(2)}_{FL} [T, V, \mu] = ~\Big[ \det \Big(   \frac{1 }{2 G_S}  -  \Pi_{\sigma}\left(q_0, \vec{q} \right) \Big)\Big]^{-\frac{1}{2}} \Big[\det  \Big(  \frac{ 1}{2 G_S} - \Pi_{\vec{\pi}} \left(q_0, \vec{q} \right) \Big)\Big]^{-\frac{3}{2}}~
\end{eqnarray}
with the polarization loop $\Pi_M \left(q_0, \vec{q} \right)$ given explicitly by
\begin{eqnarray}
\Pi_M \left(q_0, \vec{q} \right) &=& - N_c N_f ~\sum_{s, s' = \pm 1} ~ \int \frac{d^3p}{\left(2 \pi \right)^3} ~  \frac{1 - f^-_{\Phi}\big(-s'E_k\big) - f^+_{\Phi}\big(sE_p\big) }{q_0 + s'E_k - sE_p}   \Big(1 - ss' \frac{{\mathbf p} \cdot ({\mathbf p} - {\mathbf q}) \mp m^2}{E_p E_{p-q}} \Big)~,\nonumber\\
\end{eqnarray}
One easily obtains the thermodynamic potential corresponding to the chosen meson part of the partition function up to the Gaussian order given by
\begin{eqnarray}
\Omega_M^{(2)} \left(T, \mu \right) = \frac{d_M}{2} \frac{T}{V} {\rm Tr} \ln  S^{-1}_{M}~,
\label{omega_M}
\end{eqnarray}
where the degeneracy factor $d_M$ equals $1$ for sigma mesons and $2$ for pions.

From the point of view of the polarization operators, the pseudoscalar and scalar meson masses can be defined by the condition that for $q^2 = M_M^2$ the corresponding polarization operator $\Pi_M(M_M^2)$ leads to a bound state pole in the meson correlation function \cite{Hansen:2006ee}. For mesons at rest (${\mathbf q}=0$) these conditions correspond to the Bethe-Salpeter equations
 \begin{eqnarray}
1 + 4 G N_c N_f \int \frac{d^3 p}{(2\pi)^3}
\frac{4E_p}{M_\pi^2-4 E_p^2} \left( 1- f^-_{\Phi} - f^+_{\Phi} \right) &=& 0~, 
\label{masspi}\\
1 + 4 G N_c N_f \int \frac{d^3 p}{(2\pi)^3}\frac{{\mathbf p}^2}{E_p^2} 
\frac{4E_p}{M_\sigma^2-4 E_p^2} \left(1-f^-_{\Phi} -f^+_{\Phi}\right) &=& 0~. 
\label{masssigma}
\end{eqnarray}

\begin{figure}
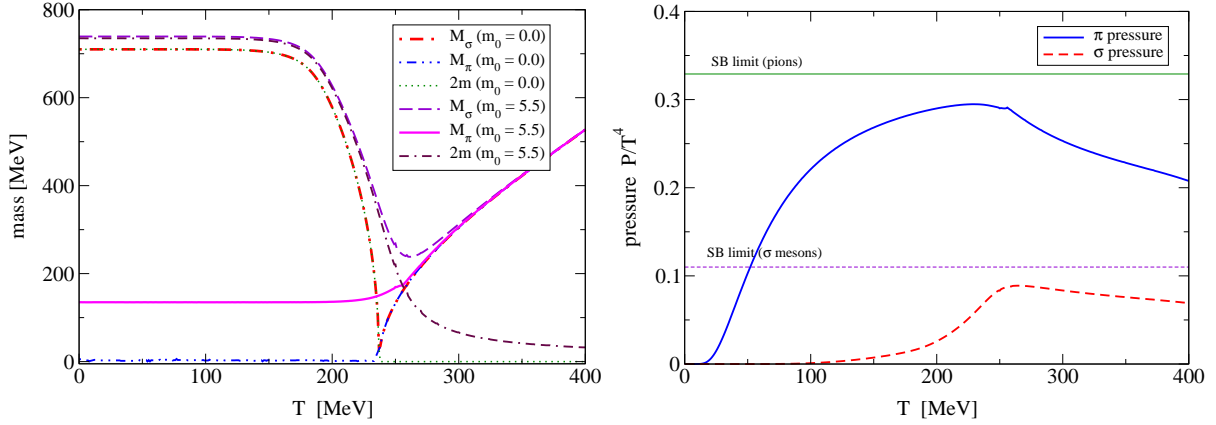

\psfig{file = 2m_Mpi_Msigma_2.eps, width = 0.48\textwidth}
\psfig{file = meson_pressure_1.eps, width = 0.48\textwidth}
\caption{Temperature dependence of the meson masses $M_M$ (left panel) and meson pressure for $\vec{\pi}$ (solid lines) and $\sigma$ (dotted lines). }
 \label{meson_pressure_87}
\end{figure}

Solutions of the two Bethe-Salpeter equations (\ref{masspi}) and (\ref{masssigma}) constitute the set of meson masses and are presented on Fig. \ref{meson_pressure_87} (left panel). The Mott temperature, which is defined by the condition $M_\pi (T_{\rm Mott})= 2m_q(T_{\rm Mott})$, is for given parameters $T_{\rm Mott} \simeq 231$ MeV in the chiral limit and $T_{\rm Mott} \simeq 256$ MeV away from it. The modification of quasiparticle properties is clearly visible: up to the Mott temperature $T_{\rm Mott}$, the $\sigma$ mass practically follows the behaviour of $2m_q(T)$, with a drop towards the pion mass, signalling chiral symmetry restoration (in the chiral limit the $\sigma$ mass exactly coincides with twice the quark mass up to $T_{\rm Mott}$). In the same region, the pion mass remains practically constant (and equal zero in the chiral limit). At $T \simeq T_{\rm Mott}$ however, the masses of chiral partners become approximately degenerate, $M_\sigma \approx M_\pi$, and then both masses increase linearly with temperature. 

The corresponding meson pressure is shown on Fig. \ref{meson_pressure_87} (right panel).

\section{Generalized Beth-Uhlenbeck approach to the quark-meson plasma}

Deriving the thermodynamic equation of state for a quark-meson system in the Beth and Uhlenbeck form reduces to the introduction of scattering phase shifts into the formula for the thermodynamic potential. This requires that we analytically continue the propagator into the complex plane.

In a first step, we introduce meson spectral functions $A_M^g (\omega, \vec{q})$ by utilizing the integral representation of the logarithm in (\ref{omega_M}), followed by the usual expression of the propagator by means of  $A_M^g (\omega, \vec{q})$ \cite{Abuki:2006dv}, i.e.,
\begin{eqnarray}
\ln  S^{-1}_{M} = - \int_0^{G_S} dg \frac{1}{2g^2} \frac{1}{\frac{1}{2g} -\Pi_M(q_0, \vec{q})} = -\int_{- \infty}^{+ \infty} \frac{d\omega}{2 \pi} \frac{1}{q_0 - \omega} \int_0^{G_S} \frac{dg}{2g^2} A_M^g (\omega, \vec{q}).
\label{interchange}
\end{eqnarray}
Using the fact that the spectral density is given by the discontinuity of the propagator at the real axis we arrive at
\begin{eqnarray}
\int_0^{G_S} \frac{dg}{2g^2} A_M^g (\omega, \vec{q}) &=& 
- i \int_0^{G_S} \frac{dg}{2g^2} 
\bigg(S_M^g\left(\omega + i\eta, \vec{q} \right) 
- S_M^g \left(\omega - i \eta, \vec{q} \right) \bigg) \nonumber\\
&=& -i\ln\left( \frac{1-2G_S \Pi_M \left(\omega - i \eta, \vec{q} \right)}
{1 -  2G_S \Pi_M \left(\omega + i \eta, \vec{q} \right)}    \right)~,
\label{to_substitute_lala}
\end{eqnarray}
where the argument of the logarithm is by definition the scattering matrix $\mathds{S}_M(\omega, \vec{q})$ in the Jost representation \cite{Hufner:1994ma}. This normalized complex function can also be represented by means of a scattering phase shift $\mathds{S}_M(\omega, \vec{q}) = \exp [2i\Phi_M(\omega,\vec{q})]$, which allows us to identify
\begin{eqnarray}
\int_0^{G_S} \frac{dg}{2g^2} A_M^g (\omega, \vec{q}) = - i \ln \mathds{S}_M(\omega, \vec{q}) = 2\Phi_M(\omega, \vec{q})~.
\label{phi_as_log}
\end{eqnarray}
Utilizing the above identity leads, after performing the Trace operation, to the following result
\begin{eqnarray}
\Omega_M^{(2)} \left(T, \mu \right) &=&  
- \frac{N_M}{2} \int \frac{d^3q}{(2 \pi)^3} \bigg(\int_{0}^{+\infty} 
\frac{d\omega}{ \pi}  
\bigg[\omega + 2 T \ln \big(1 - e^{-\beta\omega} \big)\bigg] 
\frac{d \Phi_M(\omega, \vec{q})}{d\omega} \bigg) 
\nonumber \\
&=& - \frac{N_M}{2} \int \frac{d^3q}{(2 \pi)^3} \bigg(\int_{-q^2}^{+\infty} 
\frac{ds}{\pi}  
\bigg[\sqrt{q^2+s} + 2 T \ln \big(1 - e^{-\beta\sqrt{q^2+s}} \big)\bigg] 
\frac{d \Phi_M(s)}{ds} \bigg) 
\nonumber \\
&=& - \frac{N_M}{2} \int \frac{d^3q}{(2 \pi)^3} \bigg(\int_{-q^2}^{+\infty} 
{ds}  
\bigg[\sqrt{q^2+s} + 2 T \ln \big(1 - e^{-\beta\sqrt{q^2+s}} \big)\bigg] 
D_M(s) \bigg)~,
\label{final_form_mesonic_potential_01}
\end{eqnarray}
where $D_M(s)$ is a generalized mass distribution (density of states) containing all the dynamics of the system. The above expression is the so-called generalized Beth-Uhlenbeck form of the thermodynamic potential. The connection to the original Beth-Uhlenbeck expression for the second virial coefficient can be found in \cite{Hufner:1994ma} in detail. 

For an analysis of meson masses and corresponding widths, necessary to obtain a description of mesonic correlations as bound states dissolving into resonant states above the Mott temperature, one splits the polarization function $\Pi_M(q_0, \vec{q})$ according to
\begin{eqnarray}
\Pi_M \left(q_0, \vec{0} \right) = 4 N_c N_f ~ I_1 - 2 N_c N_f P_M ~ I_2(q_0) = \widetilde{I}_1 - P_M \widetilde{I}_2(q_0)
\label{Pi_as_two_integrals_13},
\end{eqnarray}
where, in the limit $\vec{q} = 0$ that we utilize, the integrals $I_1$ and $I_2$ are given by
\begin{eqnarray}
I_1 = \int  \frac{d^3p}{\left(2 \pi \right)^3} \Bigg[ \frac{1}{2E_p}\bigg( 1 - f^-_{\Phi}(E_p) - f^+_{\Phi}(E_p\bigg) \Bigg]~,
\end{eqnarray}
\begin{eqnarray}
I_2(q_0) = \int \frac{d^3p}{\left(2 \pi \right)^3} ~ \frac{1}{4E_p^2} ~   \Bigg[ \frac{1 - f^-_{\Phi}(E_p) - f^+_{\Phi}(E_p)}{ 2E_p - q_0 }   + \frac{1 - f^-_{\Phi}(E_p) - f^+_{\Phi}(E_p)}{2E_p + q_0}   \Bigg]
~,
\end{eqnarray}
with $P_M = - q_0^2$ for pions and $P_M = - q_0^2 + 4m^2$ for sigma mesons. By identifying  $q_0 = M_M - i \frac{1}{2}\Gamma_M$ one can perform the complex mass pole analysis leading to 
\begin{eqnarray}
&& P_M = \hspace{3pt}  - \frac{\frac{1}{4N_c N_f G_S} - 2 I_1}{|I_2 (q_0 = M_M - i\frac{1}{2}\Gamma_M)|^2} \Big({\rm Re} ~ I_2 (M_M) - i ~ {\rm Im} \hspace{2pt} I_2 (M_M) \Big)~,
\label{conditions_Bethe_456}
\end{eqnarray}
which decomposes into coupled Bethe-Salpeter equations for meson mass and meson spectral width.

In the first departure beyond the pole approximation (Appendix \ref{appB456}), it is justified above $T_{\rm Mott}$ to consider $D_M(s)$ to be described by a Breit-Wigner type function
\begin{eqnarray}
A_R (s,T) = a_R \frac{M_M\Gamma_M}{\big(s - M_M^2 \big)^2 + (M_M\Gamma_M)^2 }~,
\label{BW_func}
\end{eqnarray}
where $M_M$ is the meson pole mass, $\Gamma_M$ is the corresponding meson width and $a_R$ is a normalization factor. Below $T_{\rm Mott}$, where the spectral broadening $\Gamma(T)$ of the states vanishes, the above expression becomes the delta function typical for the spectral function of a mesonic bound state. The meson phase shift $\Phi_M$ corresponding to (\ref{BW_func}) should be of the form
\begin{eqnarray}
\Phi_M (s) \approx \phi_R (s) = \frac{\pi}{\frac{\pi}{2} - \arctan\left(\frac{4m^2 - M_M^2}{M_M\Gamma_M} \right)} \Bigg(\arctan\Big[\frac{s - M_M^2}{M_M \Gamma_M}\Big] - \arctan\Big[\frac{4m^2 - M_M^2}{M_M \Gamma_M}\Big] \Bigg)~,
\label{resonant_phase} 
\end{eqnarray}
where by introducing the notation $\phi_R(s)$ we acknowledge the fact that the above phase shift is connected with resonant properties of the mesonic correlations.

\section{Levinson's theorem for quark-meson thermodynamics}

In order to inspect the validity of the approach so far presented, in our analysis we consider the Levinson theorem
\begin{eqnarray}
\int_{4m^2}^{+\infty} ds ~ \frac{d\Phi_M}{ds} = n \pi~,
\label{Levinson_987}
\end{eqnarray} 
where $n$ denotes the number of bound states below the threshold $4m^2$. Indeed, it is easy to check that the resonant phase shift $\phi_R (s)$ alone does not fulfill (\ref{Levinson_987}). This implies that the scattering phase shift should be composed of at least two parts. In fact, as was demonstrated in \cite{Zhuang:1994dw}, it is appropriate to decompose the scattering phase shift $\Phi_M$ into a part corresponding to the mesonic correlation and a part describing quark-antiquark scattering,
\begin{eqnarray}
\Phi_M = \phi_R + \phi_{sc}~.
\label{phases_876}
\end{eqnarray}
Namely, using (\ref{phi_as_log}) we can represent the total scattering phase shift $\Phi_M$ as
\begin{eqnarray}
\Phi_M = \frac{i}{2} \ln \frac{1 - 2 G_S \Pi_M(\omega + i \eta,\vec{q})}{1 - 2 G_S \Pi_M(\omega - i \eta,\vec{q})}~.
\end{eqnarray}
Then it is straightforward, using (\ref{Pi_as_two_integrals_13}) and the relation between logarithm and arctan functions, to show that
\begin{eqnarray}
\Phi_M = - \arctan \bigg[\frac{2G_S P_M {\rm Im} \widetilde{I}_2}{1 - 2 G_S \widetilde{I}_1 + 2G_S P_M {\rm Re} ~ \widetilde{I}_2} \bigg]
\end{eqnarray}
and, by several more consecutive steps, to obtain
\begin{eqnarray}
\Phi_M = - \arctan \bigg[ \frac{ \frac{{\rm Im} \widetilde{I}_2}{{\rm Re} \widetilde{I}_2} - \frac{1 -2G_S \widetilde{I}_1}{2G_S |\widetilde{I}_2|^2} \cdot \frac{ {\rm Im} \widetilde{I}_2}{P_M  + \frac{1 - 2G_S\widetilde{I}_1}{2G_S |\widetilde{I}_2|^2} {\rm Re} \widetilde{I}_2} }{1 + \frac{1 - 2G_S\widetilde{I}_1}{2G_S |\widetilde{I}_2|^2} \cdot \frac{ {\rm Im} \widetilde{I}_2^2}{P_M {\rm Re} \hspace{1pt} \widetilde{I}_2 + \frac{1 - 2G_S\widetilde{I}_1}{2G_S |\widetilde{I}_2|^2} {\rm Re} \widetilde{I}_2^2} } \bigg]~.
\end{eqnarray}
At this point it is enough to recognize the above expression as the sum of arctans to finally obtain
\begin{eqnarray}
\Phi_M = - \arctan\Big(\frac{{\rm Im} \widetilde{I}_2}{{\rm Re} \widetilde{I}_2}\Big) + \arctan\Big( \frac{1 -2G_S \widetilde{I}_1}{2G_S |\widetilde{I}_2|^2} \cdot \frac{ {\rm Im} \widetilde{I}_2}{P_M  + \frac{1 - 2G_S\widetilde{I}_1}{2G_S |\widetilde{I}_2|^2} {\rm Re} \widetilde{I}_2}\Big)~,
\label{Zhuang_decomposition}
\end{eqnarray}
which proves the formula (\ref{phases_876}), where 
\begin{eqnarray}
\phi_{sc} = - \arctan\Big(\frac{{\rm Im} \widetilde{I}_2}{{\rm Re} \widetilde{I}_2}\Big)
\label{phase_scatt_123}
\end{eqnarray}
and
\begin{eqnarray}
\phi_R = \arctan\Big( \frac{1 -2G_S \widetilde{I}_1}{2G_S |\widetilde{I}_2|^2} \cdot \frac{ {\rm Im} \widetilde{I}_2}{P_M  + \frac{1 - 2G_S\widetilde{I}_1}{2G_S |\widetilde{I}_2|^2} {\rm Re} \widetilde{I}_2}\Big)~.
\label{phase_reson_123}
\end{eqnarray}
Using the conditions (\ref{conditions_Bethe_456}), in the above equation masses and widths could be identified in accordance with (\ref{resonant_phase}).

As the first part of this decomposition is independent of mesonic properties, we presume that it is connected with the scattering states' input to the thermodynamics. On the other hand, the second part of (\ref{Zhuang_decomposition}) describes solely the behavior of dissolving mesons. Thus we are assured that the correct description of mesonic correlations accounts not only for bound and resonant states' contribution, but also for the input from the scattering states.

In our analysis we will use a combined approach, where the scattering part of the phase shift $\Phi_M$ is defined according to (\ref{Zhuang_decomposition}) and the resonant part is given by a delta function below $T_{\rm Mott}$ and by the Breit-Wigner ansatz (\ref{BW_func}) beyond it. Explicitly, we take
\begin{eqnarray}
D_M(s) = \frac{1}{\pi}\frac{d \phi_M(s)}{ds} = \begin{cases} 
\delta(s-M_M^2)+\frac{1}{\pi}\frac{d}{d s}\phi_{sc}(s)~, & T<T_{\rm Mott}~, \\
 & \\
\frac{a_R}{\pi}\frac{\Gamma_M M_M}{(s-M_M^2)^2 + \Gamma_M^2 M_M^2} + \frac{1}{\pi}\frac{d}{d s} \phi_{sc}(s)~, & T>T_{\rm Mott}~. \end{cases}
\end{eqnarray}
We will in the following regard the states for positive real $s \geq 4m^2$ only.

Finally, the scattering states' contribution to the density of states is normalized and exactly compensates the contribution from the resonance (or the bound state, resp.). This accordance with the Levinson theorem confirms the validity of the presented approach.

\section{Results and discussion}

\begin{figure}
\centerline{
\psfig{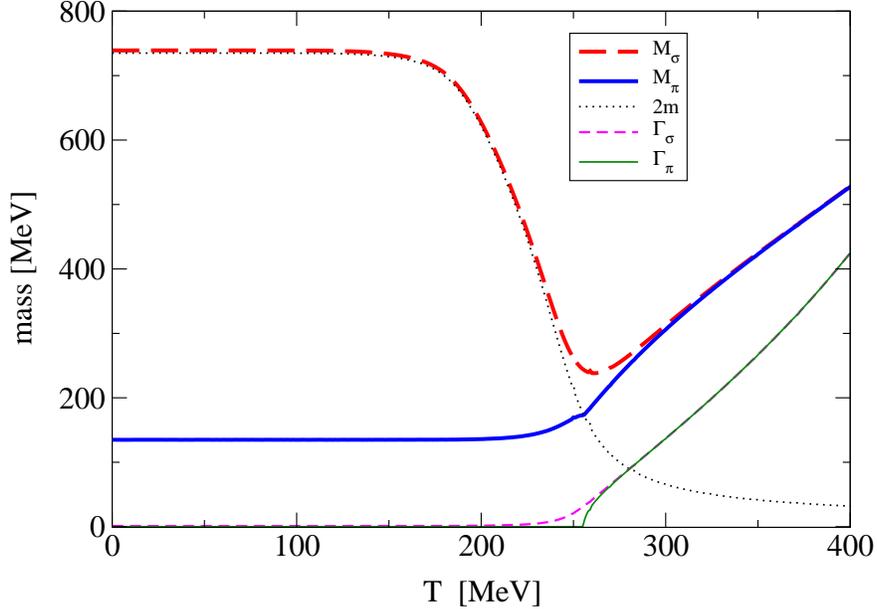}
}
\caption{ Temperature dependence of meson masses $M_M$ and corresponding spectral widths $\Gamma_M$.}
\label{masses_and_widths}
\end{figure}

The above described complex mass pole analysis leads to the results for the meson masses $M_M$ and widths $\Gamma_M$ as presented on Fig. \ref{masses_and_widths}. We observe that the sigma meson width is non-zero for all the temperatures considered, although below $T_{\rm Mott}$ it is not significant and therefore allows to consider sigma to be a quasi bound state. Above the Mott temperature pion and sigma masses quickly become equal and so do their spectral widths. This is the imprint of the chiral symmetry restoration where the $\sigma$ and $\pi$ meson, being chiral partners, become degenerate.

In what follows we will concentrate on the discussion of the $\pi$ meson, because it undergoes the Mott transition from the bound state to the resonant correlation in the continuum, accompanied by a jump of the scattering phase shift at threshold from $\pi$ to $0$ in accordance with the Levinson theorem.

\begin{figure}
\centerline{
\psfig{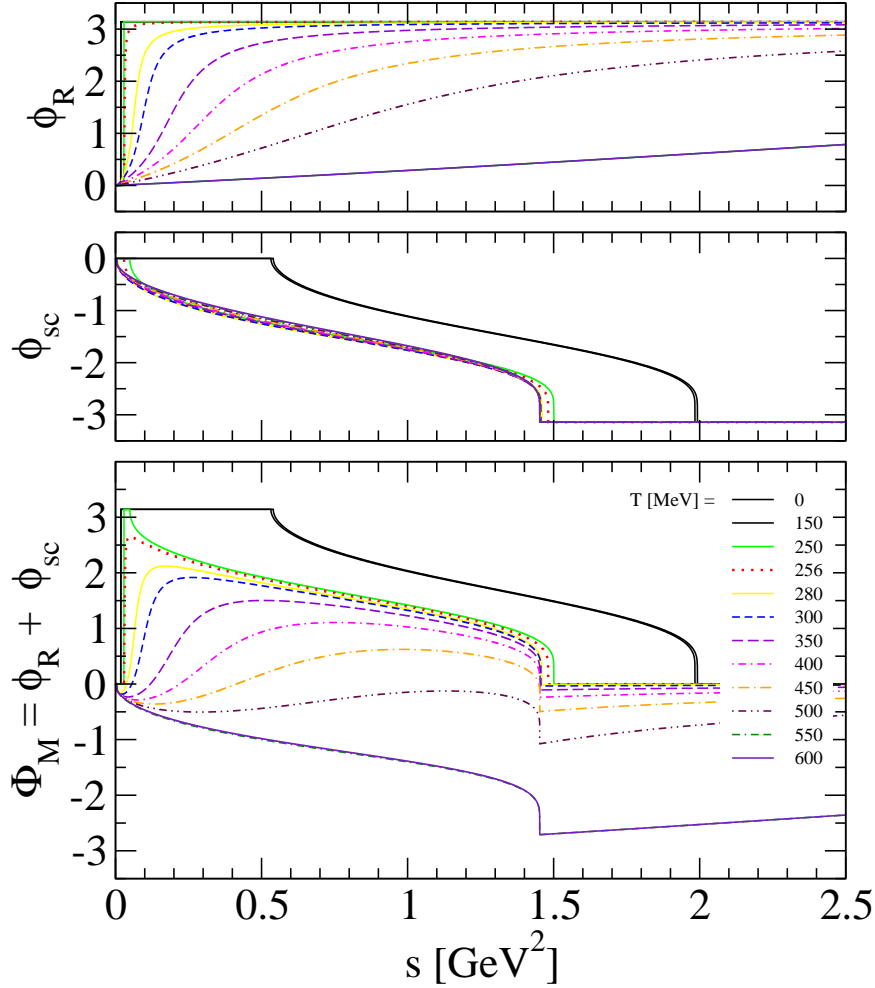}}
\caption{Dependence of phase shift $\Phi=\phi_R+\phi_{sc}$ and its components
 $\phi_R$ and $\phi_{sc}$ in the pion channel on the center of mass energy.}
\label{phase_shifts_01}
\end{figure}

In Fig. \ref{phase_shifts_01} we show the phase shift $\Phi_M$ (lower panel) in the decomposition (\ref{phases_876}) into its resonant (upper panel) and scattering continuum (middle panel) parts, obtained from the solution of equations (\ref{phase_scatt_123}) and (\ref{phase_reson_123}).

\begin{figure}
\centerline{
\psfig{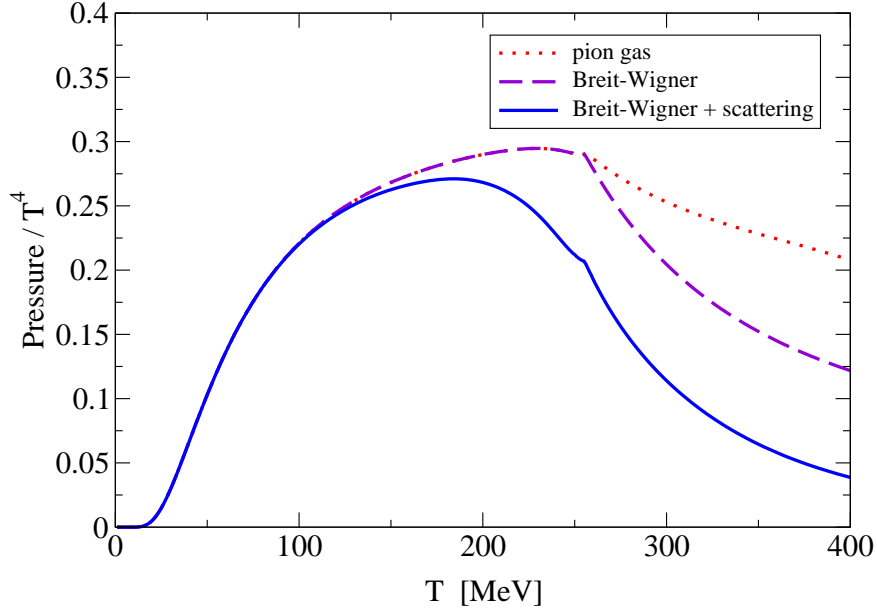}
}
\caption{Temperature dependence of the pion pressure in the approximated GBU 
approach.}
\label{pion_press_BW_phi_1}
\end{figure}

\begin{figure}
\centerline{
\psfig{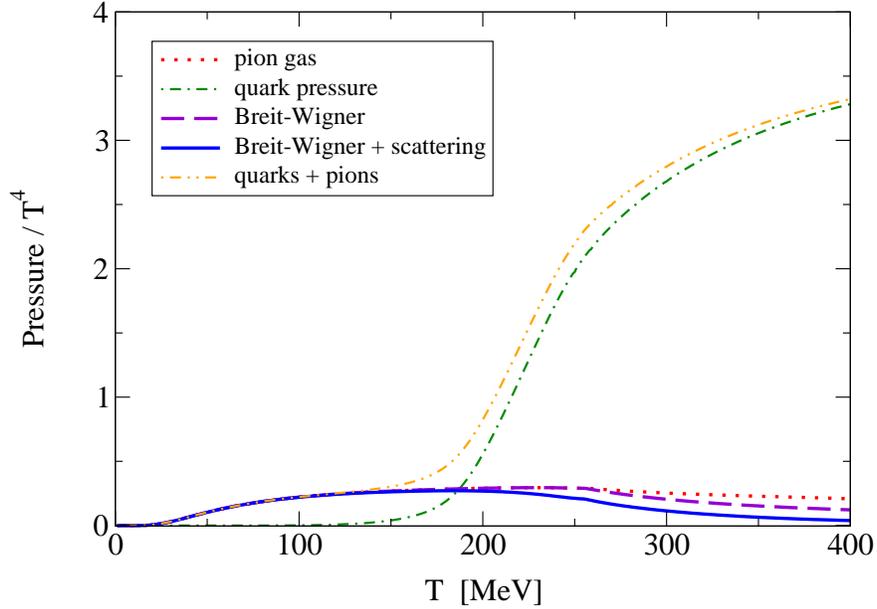}
}
\caption{Comparison of the pion pressure obtained with the input from the MFA approximation to thermodynamic potential.}
\label{pion_press_BW_phi_2}
\end{figure}
The pion pressure resulting from utilizing the Breit-Wigner spectral function 
alone and together with the input from the scattering states is shown on the Fig. \ref{pion_press_BW_phi_1} along with the pressure of the massive pion gas. Fig. \ref{pion_press_BW_phi_2} shows the pion pressure as compared with the quark pressure obtained from the mean field approximation of the PNJL thermodynamic potential. We notice that including the description of the scattering states results in a correct description of pion thermodynamics in the vicinity and beyond the Mott temperature.

\section{Conclusions}

Pion dissocation by the Mott effect in quark plasma was described
within the generalized Beth-Uhlenbeck approach on the basis of a PNJL
model whereby a unified description of bound, resonant
and scattering states was given.
As a first approximation, we utilized the Breit-Wigner ansatz for the
spectral function and clarified its relation to the complex mass-pole
solution of the pion Bethe-Salpeter equation. It has been demonstrated 
that a description of the pion Mott dissociation solely by a spectral 
broadening of the pion bound state when it enters the continuum of 
unbound states for temperatures beyond $T_{\rm Mott}$  
necessarily entails a violation of Levinson's theorem.
In order to solve this problem we have extended the approach beyond the 
complex mass pole approximation and solved the scattering phases in the
pion channel of quark-antiquark interaction. 
The account for the scattering continuum in accordance with the Levinson 
theorem leads to a strong reduction of the pion pressure above the Mott
dissociation temperature.
We suggest that the behavior of the scatterinng phase shift in the pion 
channel and its temperature dependence across the Mott transition, as obtained 
in the present work, can be used to develop a generic ansatz for the behavior 
of hadronic densities of states to be used in a generalized PNJL-hadron 
resonance gas model \cite{Turko:2011gw} that embodies the Mott dissociation 
of hadrons.

\begin{acknowledgments}
D.B. acknowledges support and hospitality during his stay at the University
of Bielefeld where this work was completed.
This work was supported in part by the Polish National Science Center (NCN) 
within the ``Maestro'' grant programme (A.W. and D.B.) under contract number 
2011/02/A/ST2/00306 and by the Russian Fund for Basic Research under grants 
No. 11-02-01538-a (D.B.) and No. 12-01-00396 (Yu.L.K.). A.W. and A.V.F. 
acknowledge support from the Bogoliubov-Infeld programme for collaboration 
between JINR Dubna and Polish Institutions.
\end{acknowledgments}

\appendix

\section{Nambu--Jona-Lasinio model with Polyakov loop}
\label{appA123}

Confinement in pure $SU(N_c)$ gauge theory can be
simulated by introducing an effective potential for a complex
Polyakov loop field. 
The PNJL Lagrangian \cite{Meisinger:2001cq}-\cite{Hansen:2006ee} is
\begin{eqnarray}
\label{Lpnjl}
{\mathcal L}_{PNJL} = \overline{q} \left(i \gamma^{\mu}D_{\mu} - m_0 - \gamma^0 \mu \right)q + \sum_{M = \sigma', \vec{\pi}'} G_M \left(\overline{q} \Gamma_M q \right)^2 - U(\Phi[A], \overline{\Phi}[A]; T)~.
\end{eqnarray}
The quark fields are coupled to the gauge field $A_\mu$ through the covariant 
derivative
$D_\mu = \partial_\mu -iA_\mu$. The gauge field is $A^\mu = \delta_0^\mu A^0 = i
\delta_4^\mu A_4$ (the Polyakov gauge). The field $\Phi$ is determined by the trace of the Polyakov loop
$L(\vec{x})$~\cite{Ratti:2005jh}
\begin{eqnarray}
\Phi[A] = \frac{1}{N_c} \mbox{Tr}_c L(\vec{x})~,
\end{eqnarray}
where $L(\vec{x}) = \mathcal{P} \exp \left[ \displaystyle -
i \int_{0}^{\beta} d \tau A_4 (\vec{x}, \tau) \right]$. $\Gamma_M$ are the vertices for the scalar ($\sigma'$) and pseudoscalar ($\vec{\pi}'$) four-fermion interaction channels. The gauge sector of the Lagrangian density (\ref{Lpnjl}) is described by an
effective potential
$\mathcal{U}\left(\Phi[A],\overline\Phi[A];T\right)$
fitted to the lattice QCD simulation results in pure $SU(3)$ gauge theory at
finite $T$~\cite{Ratti:2005jh,Roessner:2006xn} with
\begin{eqnarray}\label{effpot}
\frac{\mathcal{U}\left(\Phi,\overline\Phi;T\right)}{T^4}
&=&-\frac{b_2\left(T\right)}{2}\overline\Phi \Phi-
\frac{b_3}{6}\left(\Phi^3+ {\overline\Phi}^3\right)+
\frac{b_4}{4}\left(\overline\Phi \Phi\right)^2, \\ \label{Ueff}
b_2\left(T\right)&=&a_0+a_1\left(\frac{T_0}{T}\right)+a_2\left(\frac{T_0}{T}
\right)^2+a_3\left(\frac{T_0}{T}\right)^3~.
\end{eqnarray}

\begin{table}[!h]
\caption{Parameters of the effective potential $\mathcal{U}[A]$.}
{\begin{tabular}{|c|c|c|c|c|c|c} \toprule
$a_0$ & $a_1$ & $a_2$ & $a_3$ & $b_3$ & $b_4$ \\ \colrule
6.75 & -1.95 & 2.625 & -7.44 & 0.75 & 7.5 \\ \botrule
\end{tabular}\label{table1}}
\end{table}

The parameters of the effective potential (\ref{effpot}) and (\ref{Ueff}) are 
summarized in Table \ref{table1}.

In general, the parameter $T_0$  depends on the number of active flavors and the
chemical potential. In the present work we use $T_0 = 208$ MeV as has been proposed in \cite{Schaefer:2007pw}.

The partition function in the path-integral representation is then given by
\begin{eqnarray}
&& {\mathcal Z}_{PNJL} [T,V,\mu] = \int {\mathcal D}\overline{q} {\mathcal D}q \exp \Bigg\{ \int_0^{\beta} d\tau \int_V d^3x ~\bigg[\overline{q} \left(i \gamma^{\mu}(\partial_{\mu} - i A_{\mu}) - m_0 - \gamma^0 \mu \right)q +  \nonumber \\
&&  + ~ G_S \left(\overline{q} \Gamma_{\sigma'} q \right)^2 + G_S \left(\overline{q} \vec{\Gamma}_{\pi'} q \right)^2 - U(\Phi[A], \overline{\Phi}[A]; T)  \bigg] \Bigg\}~,
\end{eqnarray}
where the interaction vertices are written explicitly. By means of the Hubbard-Stratonovich transformation we are able to ingrate out the quark degrees of freedom to arrive at the partition function written solely in terms of collective fields
\begin{eqnarray}
&& {\mathcal Z}_{PNJL} [T,V,\mu] = \int {\mathcal D}\sigma' {\mathcal D}\vec{\pi}' \exp \Bigg\{- \bigg[ \int_0^{\beta} d\tau \int_V d^3x ~ \bigg( \frac{\sigma'^2 + \vec{\pi}'^2}{4 G_S} + U(\Phi[A]; T) \bigg) \bigg] +  {\rm Tr} \ln \Big[\beta S^{-1}[\sigma', \vec{\pi}']\Big] \Bigg\}~,
\end{eqnarray}
where $S^{-1}[\sigma', \vec{\pi}']$ is the inverse propagator given explicitly as
\begin{eqnarray}
S^{-1}[\sigma', \vec{\pi}'] = \gamma^0(i \omega_n - \mu + A_0) - \vec{\gamma}\cdot\vec{p} - m_0 - \sigma' \Gamma_{\sigma'} - \vec{\pi}' \vec{\Gamma}_{\pi'}~.
\end{eqnarray}
and the operation ${\rm Tr}$ is taken over color, flavor, Dirac and momentum indices of quark fields. 

\section{The Breit-Wigner ansatz}
\label{appB456}

We introduce the following pion propagator
\begin{eqnarray}
S_M(p^2) = \frac{1}{1 - 2G\Pi(p^2)}~.
\end{eqnarray}
In the first step we expand the polarization function around the mass pole
\begin{eqnarray}
\Pi (p^2) = \Pi(p^2 = M^2) + (p^2 -M^2)\left.\frac{d\Pi(p^2)}{dp^2} \right|_{(p^2=M^2)} + \dots
\label{pole_approx_1}
\end{eqnarray}
and we insert this expansion into the defined propagator
\begin{eqnarray}
&& S_M(p^2) = \frac{1}{\underbrace{1 - 2G\Pi(p^2=M^2)}_{= 0 ~ ({\rm by ~ definition})} - 2G(p^2 - M^2)\left.\frac{d\Pi}{dp^2}\right|_{p^2=M^2} } = -\frac{1}{ 2G(p^2 - M^2)\left.\frac{d\Pi}{dp^2}\right|_{p^2=M^2} } = \frac{g^2_{Mq\overline{q}}}{p^2 - M^2} ,
\end{eqnarray}
where in the last step we simply defined $g^{2}_{Mq\overline{q}}$. By means of the wave function renormalization we define a normalized propagator in the following way
\begin{eqnarray}
S_M(p^2) = g^2_{Mq\overline{q}} \widetilde{S}_M(p^2)~.
\end{eqnarray}
Now $\widetilde{S}_M(p^2)$ is a propagator for renormalized mass fields $\widetilde{\phi} = g_{Mq\overline{q}} \phi$. At this point we consider a complex mass pole solution
\begin{eqnarray}
p^2 = (M \pm i\frac{\Gamma}{2})^2~,
\end{eqnarray}
where for small $\Gamma$ we get $p^2 \approx M^2 \pm iM\Gamma$ and the corresponding propagator, in a similar way,
\begin{eqnarray}
\widetilde{S}_M(p^2) = \frac{1}{p^2 - M^2 \mp iM\Gamma} = \frac{p^2 - M^2 \pm iM\Gamma}{(p^2 - M^2)^2 + (M\Gamma)^2}~.
\end{eqnarray}
We introduce the spectral function by its definition to get
\begin{eqnarray}
A(s=p^2) = 2i ~{\rm Im}\widetilde{S}_M \sim \pm \frac{M\Gamma}{(s - M^2)^2 + (M\Gamma)^2}~.
\end{eqnarray}

\end{document}